# An Edge-Cloud based Reference Architecture to support cognitive solutions in the Process Industry

Antonio Salis[1], Angelo Marguglio[2], Gabriele De Luca[3], Sergio Gusmeroli[4], Silvia Razzetti[4]

[1]Engineering Ingegneria Informatica S.p.A.
Loc. Sa Illetta, SS195 km 2,3 – I-09123 Cagliari, Italy
{antonio.salis}@eng.it

[2]Engineering Ingegneria Informatica S.p.A.
Viale della Regione Siciliana Nord Ovest, 7275, I-90146, Palermo, Italy
{angelo.marguglio}@eng.it

[3]Engineering Ingegneria Informatica S.p.A.
Via Monteroni s.n., C/O Edificio Dhitech - Ecotekne, I-73100, Lecce, Italy
{gabriele.deluca}@eng.it

[4]Politecnico di Milano
Piazza L. da Vinci, 32 I-20133, Milano, Italy
{sergio.gusmeroli; silvia.razzetti}@polimi.it

**Abstract.**
Process Industry (PI e.g. Steel, Metals, Chemicals, Cement, Asphalt, Ceramics) is one of the leading sectors of the world economy, characterized however by intense environmental impact, and very high energy consumption. In spite of a traditional low innovation pace in PI, in the recent years a strong push at worldwide level towards the dual objective of improving the efficiency of plants and the quality of products, significantly reducing the consumption of electricity and $CO_2$ emissions has taken momentum. Digital Technologies (namely Smart Embedded Systems, IoT, Data, AI and Edge-to-Cloud Technologies) are enabling drivers for a Twin Digital-Green Transition, as well as foundations for human centric, safe, comfortable and inclusive work places. Currently, digital sensors in plants produce a large amount of data which in most cases constitutes just a potential and not a real value for Process Industry, often locked-in in close proprietary systems and seldomly exploited. Digital technologies, with process modelling-simulation via digital twins, can build a bridge between the physical and the virtual worlds, bringing innovation with great efficiency and drastic reduction of waste. In accordance with the guidelines of Industrie 4.0 [1], the H2020 funded CAPRI project aims to innovate the process industry, with a modular and scalable Reference Architecture, based on open source software, which can be implemented both in brownfield and greenfield scenarios. The ability to distribute processing between the edge, where the data is created, and the cloud, where the greatest computational resources are available, facilitates the development of integrated digital solutions with cognitive capabilities. The reference architecture is being validated in the asphalt, steel and pharma pilot plants, allowing the development of integrated planning solutions, with scheduling and control of the plants, optimizing the efficiency and reliability of the supply chain, and balancing energy efficiency.

**Keywords.** Industry 4.0, process industry, smart manufacturing, reference architecture, cloud and edge computing, cognitive computing, artificial intelligence, big data analytics

## 1. Introduction

Process Industry is one of the main drivers of today economy and it is characterized by extremely harsh and demanding conditions and high energy consumption. This scenario fights against an even growing pressure on environment new regulations, such as the European Green Deal initiative [2], that require enterprises to control (and reduce) their environmental impact, in terms of carbon emissions, pollutions and wastes. Additionally, the European process industry is a strong sector that needs to remain competitive in the global market, but so far it has scarcely approached innovations (plants use less advanced technologies and tools, workers and operators have poor digital background and skills). To overcome the risk of standing back to an increasing demand of process reliability and standardization of systems, that will impact the possibility of optimizing the efficiency, flexibility and time-to-market, the process industry sector requires a transformation.

Digitalization appears as the answer to the previous questions, which is going to materialize covering the entire lifecycle, from R&D to plant operations, supply chain management, customer relations and integrating material flows in a circular economy and across industry sectors [3]. Each stage of the production process may be impacted by the digital transformation, from the use of Digital Twins and predictive simulation models in the design phase, to process control and operations as well as plant reliability and regulation of maintenance activities thanks to AI and machine learning models [4, 5]. Digitalization is expected to make plants and operations eco-friendly, with an efficient energy and resource management, contributing to a significant reduction in GHG emissions, thus leading the way to a climate neutral economy [6]. Digital technologies are going to be applied pervasively in various stages of product and process development, accelerating the innovation and leading to more efficient and faster idea-to-market process, but also fostering the integration of renewable energy carriers, enhancing flexibility and diversity of energy and resource inputs, innovative materials and new business models, improving the competitiveness of the European process industry for the next decades.

The starting point for succeeding in the digital transformation is data, which derives from the large number of digital sensors that modern industries typically have installed in their plants, and the overcoming of the siloed approach based on a wide collection of data, without exploiting its great potential [7]. Keeping this in mind, it doesn't surprise that research activities are currently driven by the need of finding new solutions that enable data exploitation.

In the complex context of process industry, a number of different possible constraint and applications must be evaluated while designing any system for managing data:

- Interoperability with common industrial software is a key factor to exploit plants' data;
- Standards interfaces and protocols are fundamental to ingest data deriving from the plant;
- Accessing data, the user protection and privacy must be guaranteed;
- In process industry, the need of simulating physical models is making the Digital Twin more and more popular and it requires a specific system supporting it;
- Data-driven decisions are very likely accompanied by business intelligence tools, operational dashboards, complex event processing, machine learning algorithms;

- Cognitive models and complex computations, often based on real-time exchange of data, require high performance and speed;
- In case of data exchange with third parties, sovereignty principles must be guaranteed.

As reflected in various initiatives such as Industry 4.0 or SRIDA for the AI, Data and Robotics Partnership [8], new intelligent, robust, and scalable solutions are needed to unlock the untapped capital of data. The current paper aims at presenting a possible solution, whose adoption by process industries may represent a starting point toward digital transformation.

Within the EU H2020 funded **CAPRI Project** [9] a specific Reference Architecture has been designed and implemented to be used in the selected domains to address the cited topics. This reference architecture, distributing the processing between the edge and the cloud, can facilitate the development of integrated digital solutions embedding cognitive capabilities, thus improving production processes.

This paper is structured as follows. Section 2 introduces the CAPRI Reference Architecture, then Section 3 describes the related implementation in three different domains with status of work, with preliminary findings and expected benefits. Finally, Section 4 describes future work, contributions and importance and concludes the paper.

## 2. CAPRI Reference Architecture

CAPRI is a H2020 project that brings cognitive solutions to the Process Industry by developing, testing and experimenting an innovative Cognitive Automation Platform (CAP) towards the Digital Transformation. To achieve that, CAPRI enables cognitive tools that provide existing process industries flexibility of operation, improving the performance and state of the art quality control of its products and intermediate flows.

The CAP incorporates the methods and tools of the six Digital Transformation pathways (6P -> Product, Process, Platform, Performance, People, Partnership), engaging the cognitive human-machine interaction [10] (industrial IoT connections, smart events processing, knowledge data models and AI-based decision support). The CAP Platform and its cognitive toolbox of solutions can be replicable in areas of production planning, control, automated processes and operations of different process industry sectors.

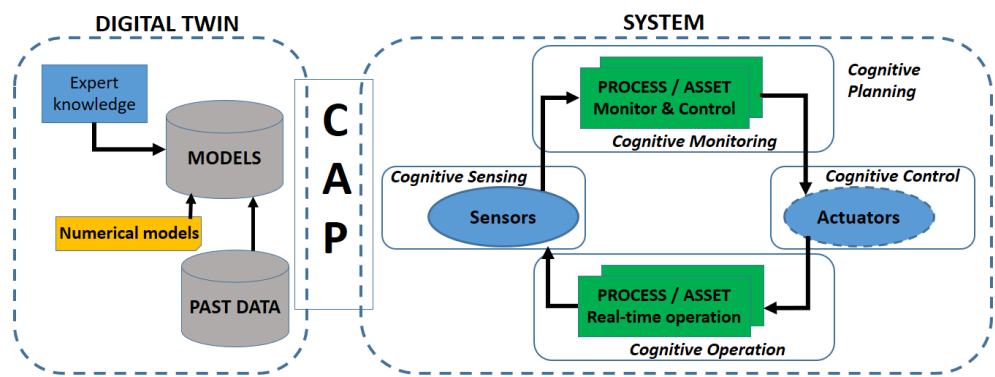

*Figure 1 - Cognition concept into CAPRI Project*

The project aims to demonstrate its applicability in the key process industries of Asphalt mix production, Pharmaceutical tablets and Steel Billets & Bars, dealing with all aspects of the interoperability between real and digital world. It will address the specific needs for process industry to work with digital twins and taking advantage from them using cognitive solutions to improve the efficiency of real plants. The project foresees the development and testing of different cognitive solutions at each automation level, from sensors, control, operation and planning. CAPRI project approach to Digital Twins is to consider them as a cognitive solution on top of lower level cognitive solutions such as cognitive sensors or cognitive control algorithms.

At the core of the CAP is the CAPRI Reference Architecture, which is based on world class open source frameworks such as FIWARE [11] and Apache [12], being compliant with state-of-the-art industrial frameworks like RAMI4.0 [13], IIRA [14], IDSA [15], BDVA [16], that enable to fit all technical challenges on data management, protection, processing, analytics, visualization, and a Data Space Enabler to get data from a highly heterogeneous set of sources and transform them into generic and configurable industrial data spaces for the process industry. The different sources support several communication protocols such as MQTT, CoAP, OPC UA, that facilitate the development of smart applications for all production processes. The Cognitive Reference Architecture will provide the glue that integrates business intelligence, operational dashboards, complex event processing, machine learning algorithms, managing the security of the data and systems and guaranteeing the data sovereignty integrating dedicated components compliant to the IDS RAM [17]. The Cognitive Reference Architecture will be the reference point for the digital transformation of cognitive plants starting from the collection of data, moving through the reasoning of the data, up to the exploitation of data with configurable dashboard. While FIWARE for Industry can provide reference implementations to most of the functional components, some improvements need to be developed to provide cognitive functions at sensor, edge and cloud level. For this aim the FIWARE for Industry RA, will be enhanced in CAPRI integrating major Apache components, by adding cognitive functions to compose the powered by FIWARE Cognitive Process Plants Blueprint, and making easier the development of cognitive components as well as the integration of them in a more comprehensive way, so the ability to approach both a greenfield and a brownfield.

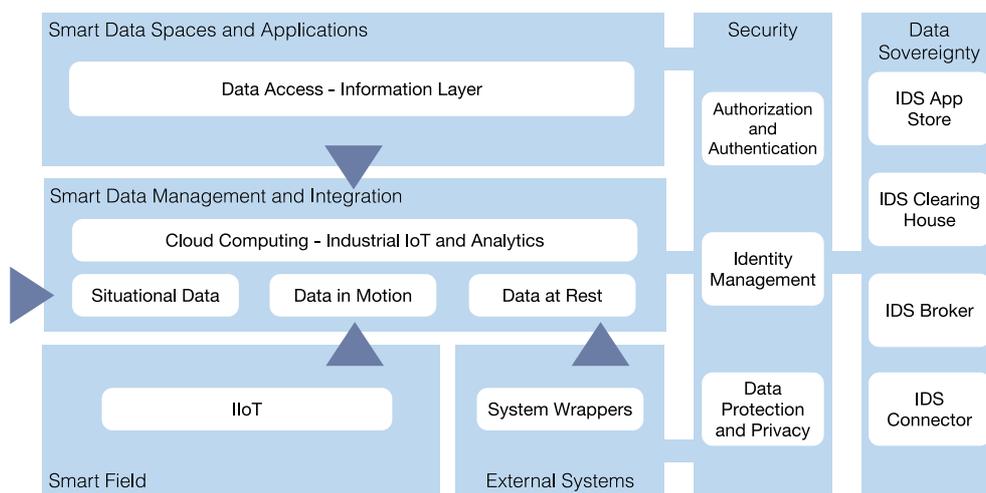

*Figure 2 - Three-tiers CAPRI Reference Architecture*

The resulting CAPRI RA and Blueprint will enhance the capability to reach the expected benefits and impacts on the proposed use case, but also to apply many of the cognitive solutions to other sectors. From the Asphalt domain, some solutions could be applied to process plants in cement and mineral sectors. Steel production domain has similarities with copper, aluminum, zinc, and other metal-processing industries. For both cases final modelling is different and must be tailored case-by-case, but the overall method, sensorial equipment, training and algorithms can be directly transferred, and this widens impact, making possible the exploitation into different markets and sectors. From the Pharma domain, the resulting solutions could be shifted to any other industry based on processing of powder raw material. The investigated unit operations feeding, granulation and fluid bed drying are widespread in industry. All sectors utilizing the mentioned process steps can benefit from implementation of the concept of in-line monitoring of attributes like concentration, blend uniformity and granule size distribution, the modelling of mentioned unit operations and the development of the cognitive control concept. Food industry is especially related to pharma processing and faces the same challenges. The concepts developed for the tableting process can be employed in future for many consumer products such as dishwashing tabs, etc.

### 2.1 The Reference Architecture layering

The resulting three-tier RA is depicted in Figure 3 and defines several functional macro-components:

- **Smart Field** represents the physical layer and contains industrial devices, machines, actuators, sensors, wearable devices, robots, etc. that are spread in the shop floor, and supports the most common industrial and, more in general, IoT protocols such as OPC UA, MQTT, etc. Standards interfaces and protocols must be used, in order to represent the information collected from the plant and to connect and integrate actuators for implementing the sensing and control mechanisms. Data will be collected typically as Data in Motion (DiM) since data coming from IIoT systems are dynamic and should be ingested and processed in real time.
- **External Systems** component contains all internal and IT systems for supporting industrial processes (ERPs, PLMs, Supply Chain Management, customized, etc.). It represents static information that comes from Legacy Systems and can be collected as Data at Rest (DaR). Custom interfaces and system wrappers are a crucial part of the component, aiming to share data using smart data models for representing information.
- **Smart Data Management and Integration** is the core of the architecture since it contains the brokering, the storage and the data processing capabilities, including cognitive process analytics and simulation systems. Data in Motion (DiM), Data at Rest (DaR) and Situational Data are represented using standard information models and made available using standard APIs. Thought the service layer, data can be collected and persisted supporting a wide range of database (i.e., relational, NoSQL, time-series).

    Data Ingestion sublayer provides a bridge between the physical layer and the data brokering, where the data from the devices are shared in a standardized structure with the broker, putting the information at the disposal of the tools will analyse them. FIWARE IDAS Generic Enabler is the IoT component that translates IoT-specific protocols into the NGSI-LD context information protocol, which is the FIWARE standard data exchange model. IoT Agent for OPC UA, IoT Agent for JSON, IoT Agent for Ultralight are some IDAS Agents in FIWARE Catalogue.

    The Data Brokering sublayer role is to manage the persistence and processing phase, where the main actors are the Orion-LD Context Broker, able to manage the entire

lifecycle of context information including updates, queries, registrations, and subscriptions and Apache Kafka for high-performance data pipelines, streaming analytics, data integration, and mission-critical applications.

The Data Persistence and Processing sublayer is composed of various FIWARE (Cygnus, Quantum Leap. Draco, Cosmos) and Apache (Livy, Spark, StreamPipes) components and is devoted to storing the data collected and process them. Cygnus, Quantum Leap and Draco are in charge to support the data storage (and pre-processing) acting as a data sink for the persistence vertical. Spark is a parallel processing framework for running largescale, both batch and real-time, data analytics applications across clustered computers. Data flows can be defined with Draco running Spark jobs through Apache Livy. StreamPipes is an Industrial IoT toolbox to enable non-technical users to connect, analyze, and explore IoT data streams. Its runtime layer supports the addition of pipeline elements through a built-in SDK in the form of microservices.

The Data Visualization gives a clear understanding of resulting data giving it visual context through maps or graphs. There are specific components, compliant with the most data source that fit different scenarios: Wirecloud enable the quick creation of web applications and dashboards/cockpits, while Grafana supports the analytics and interactive visualization, more oriented to complex monitoring dashboards. Knowage offers complete set of tools for analytics, paying attention in particular at the data visualization for the most common data sources and big data, covering different topics like Smart Intelligence, Enterprise Reporting, Location Intelligence, Performance Management, Predictive Analysis. Finally, Apache Superset is fast, lightweight, intuitive, and loaded with options that make it easy for users of all skill sets to explore and visualize their data, from simple line charts to highly detailed geospatial charts.

- **Smart Data Spaces and Applications** represents the data application services for representing and consuming historical, streaming and processed data. A wide spectrum of domains and class of applications are supported:
    o BI & Analytics increasing the business value supporting augmented intelligence and machine learning for implementing data-driven and cognitive decision-making.
    o AR/VR offering services for supporting the decision-making process (improve the human-machine interactions, accomplish proficient operational intelligence, etc.).
    o Chatbots & Virtual Assistants to enrich (chatbots) and assists (virtual assistants) the users (blue-collars and customers).
    o Novel HMI implementing new user experience developments such as supporting inexperienced operators, machine-human-machine operations, operator decision-making, etc.
    o Self-service Visualization supporting business users for accessing all data features and make data-driven decisions in a quick and scalable way.
    o Generic Cognitive Applications, implementing the cognitive manufacturing such as the self-learning, the continuous learning, the machine reasoning, the communication in natural language.
- **Persistence** represents a vertical aiming to store and make data available for the rest of the architecture. It supports a widespread of databases from Hadoop to the classic relational, passing through the NoSQL ones.
- **Security** defines components for the authorization and authentication of users and systems, integrating modules for data protection and privacy. Keyrock is the main component for Identity Management, which provides OAuth2-based [18] authentication and authorization security to services and applications, calling the AuthZForce API to get authorization decisions based on authorization policies, and authorization requests from PEPs.
- **Data Sovereignty** contains the components of the IDS ecosystem able to exchange data in a secure way guaranteeing the technological usage control and the implementation of the data sovereignty principles. The True Connector plays a major role as it is a technical component, based on IDS standards, to standardize data exchange between participants in the data space.

The CAP Reference Architecture aims to cover several industrial scenarios, from the edge to the cloud processing, passing through the Big Data analytics. In that sense, supporting a wide spectrum of databases is not sufficient, advanced processing capabilities must be supported. The RA is able to support the analysis of streaming and batch data acquired from heterogeneous external sources with the support of machine learning technologies. In particular, a typical scenario in the IoT and industrial field is to react to events in real time based on the knowledge of past events, an essential information for predicting future behavior, for example in order to identify any anomalies. In that sense, the platform integrates cognitive computing services in order to learn from experience and derive insights to unlock the value of big data. **Figure 3.** shows the final implementation.

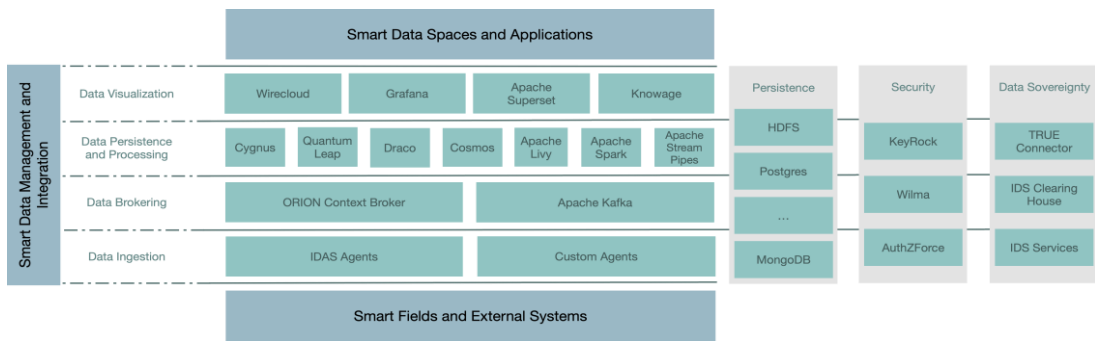

*Figure 3 - CAP Final Implementation*

### 2.2 Modularity and edge-cloud balancing

The described architecture has been conceived with modularity as a main principle: components in every layer can be combined according with a Lego-like approach, fulfilling the exposed data schema, making the architecture flexible and adaptable to the specific needs of the various application domains in process industry.

At the same time the modularity makes possible to approach a microservices design of the application that produces smaller software code, to be organizer as docker containers, so they could be run on smaller processing elements and restricted resources, as we can find in current plants, thus making easier the reuse of existing computing equipment.

In this respect, the CAPRI RA allows the implementation on both cloud and edge, managing the edge-cloud continuum [19, 20, 21]: **Figure 4.** shows the edge version of the RA, that can be run on virtualized computing resources nearer to where multiple streams of data are created, thus addressing system latency, privacy, cost and resiliency challenges that a pure cloud computing approach cannot address, and make a big difference in process industry. The edge implementation smoothly integrates with the cloud version, to enable data collection, storing, processing and presentation directly from the plant. Most of the short-term processing, including some data analytics, artificial intelligence and cognitive tasks could be managed at the edge, while cloud resources can be devoted to non-mission critical - massive processing of data.

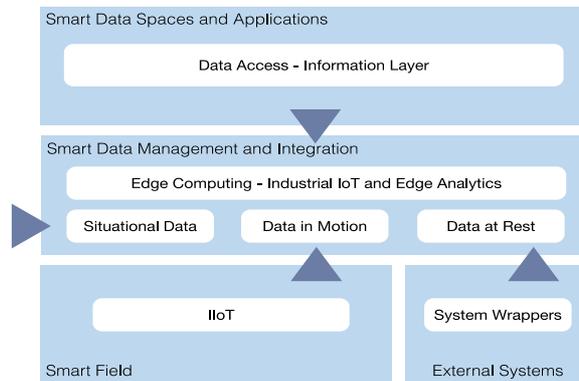

*Figure 4 - CAP Reference Architecture for Edge*

The ability to use both scenarios is very relevant in all cases where the specific application domain requires the capabilities to ingest, process and consume data with low latency and fast reactivity. Some cognitive services can be integrated and made available in the Edge Analytics module for supporting the operational intelligence and blue-collars activities in the plant.

The Smart Data Management and Integration component will act as a data provider for the analogous cloud one in order to send edge data (data in motion and/or processed information) to the cloud for longer term processing and archiving.

Summarizing the proposed architecture enables to schedule of specific components in the best possible processing elements in order to optimize the performance, network usage, fulfilling both quality and security requirements: the orchestrator and resource management elements provide the requested support.

## 2.3 Support for Cognitive solutions

One of the most relevant challenges in developing innovative solutions in the process industry is the complexity, instability and unpredictability of the processes, since they usually run in harsh condition, dynamically changing the values of process parameters, missing a consistent monitoring/measurement of some parameters important for analyzing process behavior. For cognition-based solutions these are even more critical constraints, since cognition requires a huge amount of high-quality data for ensuring the quality of the learning process in terms of precision and efficiency. Moreover, getting high quality data usually requires an intensive involvement of human experts in curating the data in a time-consuming process. In addition, a supervised learning process requires labelling/classifying the training examples by domain experts, which makes a cognitive solution quite expensive. From all these points it follows that the role of human is critical for process quality control: the expert operators have the sixth sense to early detect variations/unusualities in the process and reliably decide on-fly if the unusuality is something that should be followed closer or is just a temporary disruption.

The CAPRI RA is based on recent work in cognitive science [22], reflecting the need to explain the intelligent behavior through cognitive processes, as they are structured in human beings.

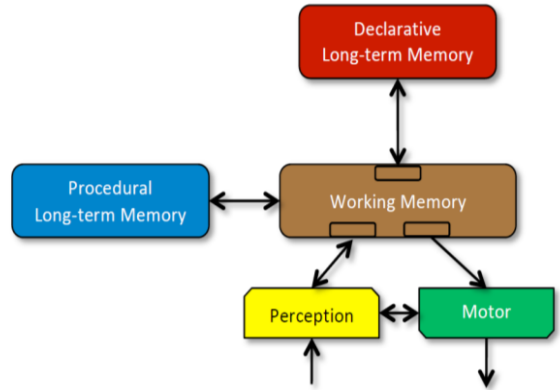

*Figure 5 - Cognitive Architecture across Artificial Intelligence, Cognitive Science, Neuroscience and Robotics*

**Figure 5.** shows the cognitive architecture and explains that human-oriented processes can be mapped into industry processes: **Cognitive Perception** can be obtained through dynamic data-driven scanning of complex industry assets, realizing an efficient, cognition-based collection of heterogeneous data, while fast Thinking enables an efficient edge-based variations detection and fast understanding of new situations, creating attention for complex, unknown situations. Slow Thinking is the efficient processing of complex situations, by creating digital models of their behaviour from sensed data, supporting timely and precise decision making, to provide a mapping between the cognitive architecture and the CAP. So in the Reference Architecture the Processing Layer, using advanced AI modules and combining them in StreamPipes, integrates the smart cognitive components and enables the cognitive perception. Data Processing and Analysis layers of the BDVA [16] reference model have been mapped and implemented by several interchangeable and open-source components from FIWARE and APACHE ecosystems. At the same time levels of cognitive human-machine interaction can be represented by the Application Layer (Data Visualization and User Interaction) and actuation in the Smart Field Layer (Edge, IoT, CPS).

## 3. CAPRI CAP in practice

The implementation of the data-driven Reference Architecture for Cognitive Plants in the process industry has followed been applied to three different domains, the key process industries of Asphalt mix production, Pharmaceutical tablets and Steel Billets & Bars. For each of these domains the standalone cognitive solutions for sensing, control, operation and planning have been mapped to the specific CAP layers and integrated in the final solution, delivering the cognitive applications towards the appropriate user role, such as planners, managers and workers. A relevant point is made by the vertical integration of layers in the CAP, so the design of the interfaces to make available the services to the upper layer, for all layers.

The development methodology follows an incremental approach and is based on two main stages: the first one is acting at the level of individual smart cognitive components, and each domain draft a schema of dependencies of the single use case to the CAPRI CAP, to shape the specific requirements, then the second stage aims to evolve the maturity of prototypes with the integration of the smart modules within the CAP to reach the objectives of quality, flexibility and performance. These two stages are functional to the mapping of the defined

CAP layers to the three different use cases involved in CAPRI to provide cognitive capabilities.

The validation follows a bottom-up approach, with the development of standalone cognitive solutions in laboratory for each single domain, then a generalization of the implementation in the CAP layers service architecture that will drive the integration phase, and the final step will be the demonstration and validation of all solutions in a real scenario in the three different sectors.

In this final stage blueprints are prepared with the aim to describe how to best use the CAP framework to implement cognitive solutions in specific industrial sectors, remaining agnostic to the specific pilot, and generic enough to represent the industrial sectors in CAPRI.

### 3.1 CAP in the Asphalt domain

The asphalt use case is based in the asphalt mix batching plant of EIFFAGE in Gerena, Seville. This plant is a mobile one, composed by nine important parts: the cold aggregates bins feeders, the dryer drum, the baghouse filter, hot aggregates bins, system of filler input, line of RAP, bitumen tanks, fuel tank and the mixer. Currently the plant has a low level of cognition, so the implementation of the CAPRI CAP in asphalt serves to improve this level, guaranteeing the optimization of processes, lower consumption of energy and reduced emission of $CO_2$.

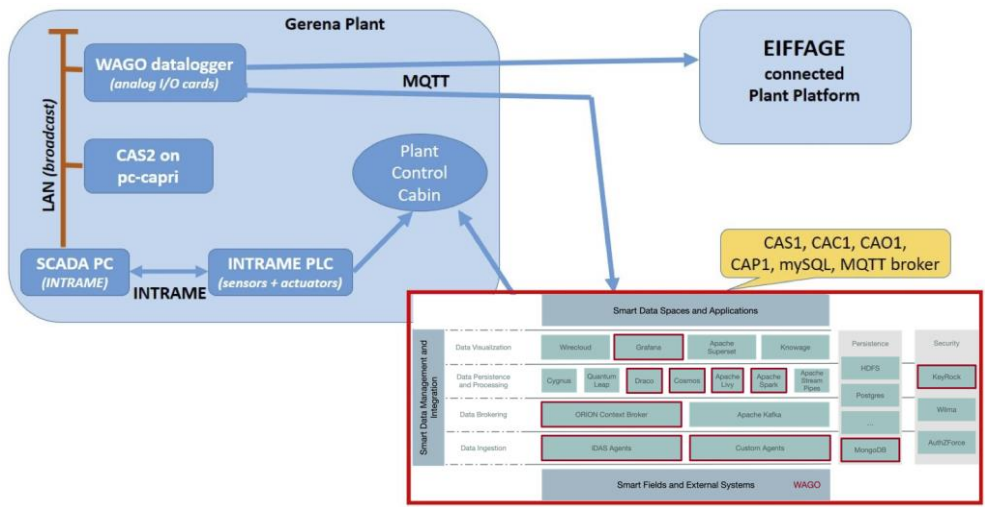

*Figure 6 - CAP Implementation in the Asphalt domain*

At the sensors level, data are collected through a direct connection to WAGO data logger, using ad hoc customized connectors or, as an alternative/parallel option, using well known standards like MQTT. Once collected, data is made persistent in Draco (a customization of Apache NiFi, passing through Orion-LD, following the NGSI-LD standards. Data processing oversees Spark, enabled by Livy, which using its features, will provide an accurate dataset to Grafana to properly visualize the desired output. Data needs to be protected due to industrial property and for GDPR compliance, this part will be covered by KeyRock. **Figure 6.** shows the Asphalt integration scenario.

### 3.2 CAP in the Steel Domain

The steel use case is based in the SIDENOR's Basauri steel plant, which has facilities for melting steel scrap, secondary metallurgical treatments, continuous casting, hot rolling and finishing of the steel products. The plant produces steel billets in the continuous casting machine as intermediate products, and some of them are further processed to steel bars in the hot rolling mills and finishing lines in the plant. The steelmaking plant has a very high demand of electrical energy, for casting of each melt of a specific steel grade, an appropriate process route needs to be chosen, according with current plant availability, technical possibility and energy prices and availability.

The implementation of the CAPRI CAP in Steel serves to help operators to make best decisions on how to manage the steel production in the plant, optimizing the planning, control and quality of produced steel bars, minimizing the consumption of energy. In the steel use case cognitive sensors, coming from PLC and SCADA systems, provide enhanced process data that serves as input for the risk estimator and offer further insights for the operator. These data are collected through the Orion-LD Context Broker (following the NGSI-LD standards) and Apache Kafka, and made persistent through Draco, driving advanced analytics processing using Apache Spark, for the optimization of production planning.

The brokering components helps also to exchange data with the existing digital twin, thus enabling simulations based on physical models, data-driven models and combinations of both, to spot and prevent process anomalies or damages in the plant, suggesting corrective actions.

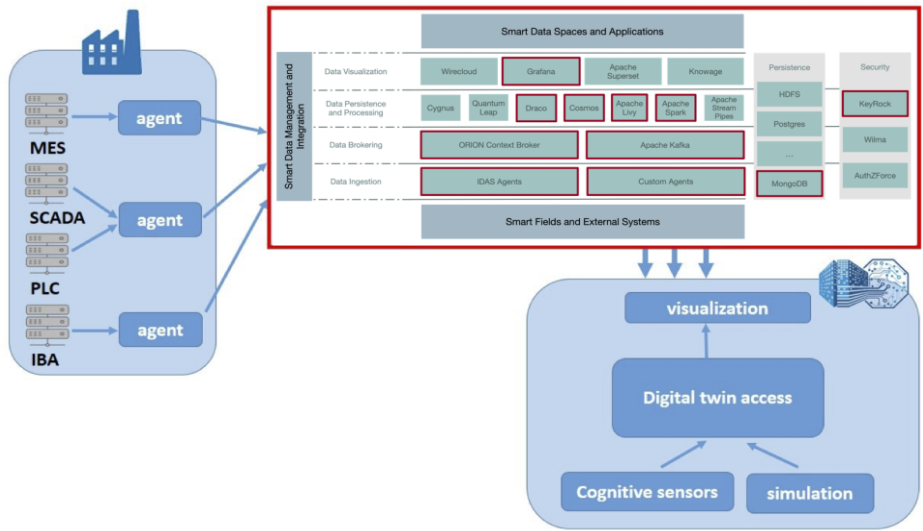

*Figure 7 - CAP Implementation in the Steel domain*

Finally, the output dataset can be properly visualized leveraging the Grafana capabilities. All aspects related to industrial property protection and GDPR compliance is managed by KeyRock. **Figure 7.** shows the Steel integration scenario.

### 3.3 CAP in the Pharma domain

The pharma use case is based in the RCPE pilot plant in Graz, supporting a continuous operating line for pharmaceutical manufacturing process of tablets, via the wet granulation route. The main unit operations are twin-screw wet granulation, followed by a dryer, a mill, a blender and a tablet press. Granules are formed using a liquid binder, then wet granules are dried in a fluid bed dryer (FBD). A mill reduces the size of larger agglomerates to the desired particle size distribution of granules before additional excipients are added. A ribbon blender guarantees blend homogeneity before material is transferred to the tablet press, while in the rotary tablet press the granule/powder blend is compacted by upper and lower punches within matrices.

The implementation of the CAPRI CAP in Pharma serves to help operators to improve real-time control strategies based on acquired process data for closed-loop process control of an entire manufacturing line.

In the Pharma use case, the data will be collected through an IDSA Agent, for example the OPC-UA Agent, that is a plug and play component used to transfer and share contexts to Orion-LD from the OPC UA servers at which it is connected. Whenever the data is collected, they will be stored in Draco. The Orion-LD broker helps in feeding collected data to processing jobs: Apache Livy supports data processing starting Apache Spark's jobs, and AI based libraries serve to predict anomalies or wastes in the production line. Spark module is able also to run MATLAB based algorithms. Finally, the output dataset can be properly visualized leveraging the Grafana capabilities. All aspects related to industrial property protection and GDPR compliance is managed by KeyRock. **Figure 8.** shows the Pharma integration scenario.

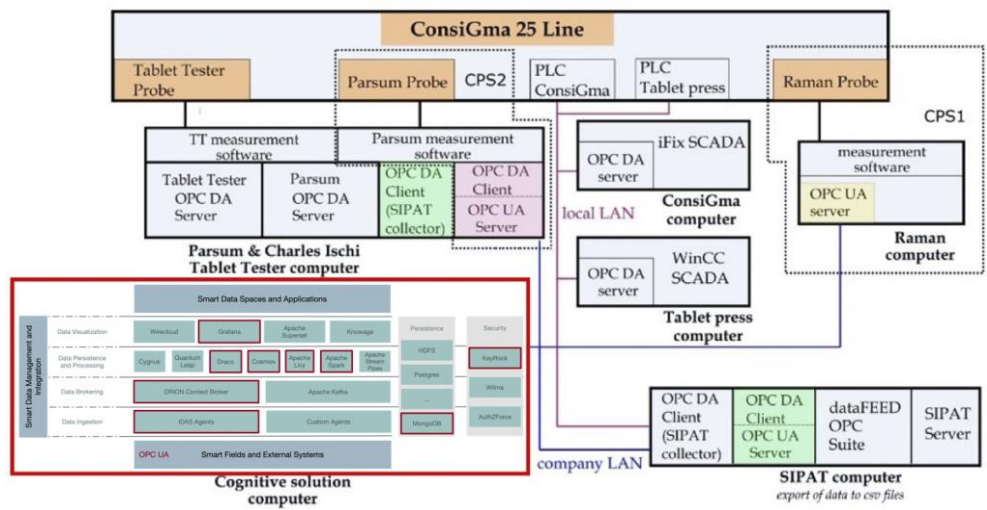

*Figure 8 - CAP Implementation in the Pharma domain*

### 3.4 Benefits and outcomes

Process Industry is a relevant industry sector with limited use of advanced equipments and where digital transformation can boost the efficiency of production while reducing the energy consumption and $CO_2$ emissions. The described CAP is a standard OSS based platform that answers the question of easing access to the provision of a complete digital toolkit with cognitive services as "plugin", for Process Industry and beyond.

From an ICT perspective the CAP offers further benefits coming from the ability to run implemented software on both the edge of the plant in existing devices as in the case of a brownfield, where data are generated and where response time could be an issue or critical factor, at the same time the software could be run in the cloud to get advantage of unlimited processing capabilities that could be needed. On top of this the CAP offers features to orchestrate the processing between the edge and the cloud, optimizing the use of available computing resources according to the business needs and objectives.

The CAP platform aims to cover the lack of companies' skills regarding the coding, lowering the barriers for implementing the digital transformation and adopting the cognitive functionalities in the process industries.

It's worth to mention that the project is part of a wide network of different stakeholders (enterprises, technology providers, etc.) where the results and experience can be shared and synergies exploited in the validation phase. Actually, CAPRI is part of the SPIRE Ecosystem [23] and specifically it is one of the six members of the SPIRE-06 cluster [24], a European Commission call grouping projects aiming at developing new technologies to realise cognitive production plants. A wider range of process industry sectors are involved (Textile, Aluminium, Cement, Oil, etc.), where the RA can be validated and adapted, if required.

Furthermore, since the other projects are pursuing the same goal (to apply the digital transformation to the full industrial process), the cluster represents a fertile ground where to start a discussion and compare the results, paving the way to further improvements and refinements. Among cluster's members, similar actions are in place and, now that the CAP has reached the maturity, a collaborative approach is foreseen: for instance, the HyperCOG project [25] has implemented an innovative Cyber-Physical System (CPS), based on an Hyperconnected Architecture [26, 27], and following current trends in I4.0 [28] that will be subject of analysis in next months. At the same time the Cognitwin project [29, 30] uses the RAMI AAS and Apache Streampipes for the design of hybrid digital twins [31, 32] that will be deeply evaluated and compared with the CAPRI approach.

## 4. Conclusions

The Reference Architecture, developed in the CAPRI project and described in previous sections, represents a fundamental enabler for the Twin Transition of process industry, becoming not just more competitive in the global marketplace but also stimulating environmental sustainability, waste reduction and circularity as well as human centricity and wellbeing., The proposed Reference Architecture includes an Open Source reference implementation conceived to materialize in concrete industrial cases the Twin Transition principles. Based on Open Standards and Data Models, CAPRI RA has been developed to have a wide applicability in the PI, both in terms of sectors and mainly in terms of available functionalities, as it supports a large variety of data applications (from simple computation and basic results visualizations to complex machine learning models and digital simulations), by encompassing a number of horizontal layers, able to guarantee interoperability, privacy, protection and data sovereignty.

The ongoing CAPRI implementation of the three use cases involving three different sectors (asphalt, steel, pharma) with the enhancement of novel cognitive functionalities allows us to collect a relevant number of lessons learnt. First of all, in a context (the process industry) usually reluctant to innovation, the Reference Architecture shows that easy-to-implement, low cost and human-centric digital solutions represent a key success factor also in PI. CAPRI RA has been conceived to provide the capabilities for the analysis of processes and to support the implementation of cognitive solutions; additionally, it overcomes the typical siloed approach, since it allows to merge and combine the information generated during several production steps.

The implementation of the use cases is a concrete example of possible applications that can be supported by the proposed RA, all of them of fundamental importance in process industry activities. For instance, the CAP platform is used in:
1. fully dynamic and model-based control of single processes;
2. decisional support for all processes and process chains where model-based online control is not yet possible;
3. coordination and optimization system for the control of interconnected processes in a process chain;
4. ecosystem for suitable digital twin of plants, processes and materials.

The next step is to leverage on the use cases to show that the implementation of the CAPRI Reference Architecture represents the enabler to develop a number of data-driven cognitive models and solutions that will be fundamental for process industry to reduce the environmental impact and to increase flexibility and quality, in accordance with the European Commission guidelines.

In particular, being CAPRI committed to improve process industry efficiency leveraging on digitalization, the Reference Architecture's return can be also measured in terms of reduction of the environmental impact and carbon footprint. To do it with CAPRI pilots, we integrated and deployed the cognitive solutions implemented on top of the CAP platform, together with the different indicators that have been identified to evaluate their footprint. Some KPIs are very specific for the sector and module developed, others are extremely generic and they can be applicable in many other contexts, as energy consumption waste reduction, maintenance costs, personnel costs, quality of products, productivity (output over consumed hours).

So far, the Cognitive Automation Platform defined in the Reference Architecture has been implemented as a first prototype, together with the cognitive solutions on top of it, and needs to be further validated in the CAPRI project's third year of activity. Next project's months will be useful to properly evaluate the pilots' performance improvement, measuring on medium term how the adoption of the Reference Architecture can impact on the productivity and energy consumption.

In a long-term perspective, the validation of the CAP will be extended outside of the CAPRI borders and the results achieved will be replicated in other sectors with similar challenges from the point of view of cognitive solutions applied to similar processes.

**Acknowledgments** This work has been supported by the CAPRI project, which has received funding from the European Union's Horizon 2020 research and innovation programme under grant agreement No. 870062.

**References**
1. Platform Industrie 4.0, https://www.plattform-i40.de/IP/Navigation/EN/Home/home.html
2. *European Green Deal,* https://ec.europa.eu/info/strategy/priorities-2019-2024/european-green-deal_en
3. T.Ritter, C.L.Pedersen, Digitization capability and the digitalization of business models in business-to-business firms: Past, present, and future. Industrial Marketing Management, Volume 86, 2020, Pages 180-190, https://doi.org/10.1016/j.indmarman.2019.11.019

4. P. Aivaliotis, K. Georgoulias, Z. Arkouli, S. Makris, Methodology for enabling Digital Twin using advanced physics-based modelling in predictive maintenance, Procedia CIRP, Volume 81, 2019, Pages 417-422, https://doi.org/10.1016/j.procir.2019.03.072
5. Stojanovic L: Cognitive Digital Twins: Challenges and Opportunities for Semantic Technologies (Keynote). SeDiT@ESWC 2020
6. OECD (2019), "Transition towards a climate-neutral economy", in Regions in Industrial Transition: Policies for People and Places, OECD Publishing, Paris. DOI: https://doi.org/10.1787/81ebdb4c-en
7. Örjan Larsson, AI & Digital Platforms: The Market [Part 1], AI and Learning Systems - Industrial Applications and Future Directions, Konstantinos Kyprianidis and Erik Dahlquist, IntechOpen, DOI: 10.5772/intechopen.93098
8. "SRIDA for the AI, Data and Robotics Partnership, Third Release." September 2020. https://bdva.eu/sites/default/files/AI-Data-Robotics-Partnership-SRIDA%20V3.1.pdf
9. http://www.capri-project.eu
10. "Digital Transformation Pathway - 6Ps methodology." 2020. https://ec.europa.eu/research/participants/documents/downloadPublic?documentIds=080166e5d71452aa&appId=PPGMS
11. "FIWARE" https://www.fiware.org
12. "Apache" https://www.apache.org
13. "RAMI4.0 - a reference framework for digitalization" https://www.plattform-i40.de/IP/Redaktion/EN/Downloads/Publikation/rami40-an-introduction.pdf?__blob=publicationFile&v=3
14. "Industrial Internet Reference Architecture" https://www.iiconsortium.org/IIRA-1.7.htm
15. "International Data Space Association (IDSA)" https://internationaldataspaces.org/
16. "Big Data Value Association/DAIRO" https://www.bdva.eu/
17. "IDS Reference Architecture Model (RAM)" https://internationaldataspaces.org/use/reference-architecture/
18. D. Hardt (ed.): RFC 6749: The OAuth2 Authorization Framework, IETF Oct. 2012, https://www.rfc-editor.org/rfc/rfc6749.txt; DOI:10.17487/RFC6749
19. Masip-Bruin X, Marín-Tordera E, Sánchez-López S, Garcia J, Jukan A, Juan Ferrer A, Queralt A, Salis A, Bartoli A, Cankar M, Cordeiro C, Jensen J, Kennedy J. Managing the Cloud Continuum: Lessons Learnt from a Real Fog-to-Cloud Deployment. *Sensors*. 2021; 21(9):2974.
20. Masip-Bruin X, Marín-Tordera E, Juan Ferrer A, Queralt A, Jukan A, Garcia J, Lezzi D, Jensen J, Cordeiro C, Leckey A, Salis A, Guilhot D, Cankar M: mF2C: towards a coordinated management of the IoT-fog-cloud continuum. SmartObjects@MobiHoc 2018: 8:1-8:8
21. Masip-Bruin X, Marín-Tordera E, Juan Ferrer A, Salis A, Kennedy J, Jensen J, Jukan A, Bartoli A, Badia R, Cankar M, Begin M: mF2C: The Evolution of Cloud Computing Towards an Open and Coordinated Ecosystem of Fogs and Clouds. Euro-Par Workshops 2019: 136-147
22. Laird, J. E., Lebiere, C., & Rosenbloom, P. S. (2017). A Standard Model of the Mind: Toward a Common Computational Framework across Artificial Intelligence, Cognitive Science, Neuroscience, and Robotics. AI Magazine, 38(4), 13-26. https://doi.org/10.1609/aimag.v38i4.2744
23. https://www.aspire2050.eu/sites/default/files/users/user85/SPIRE viewpoint_towards FP9.pdf
24. SPIRE06.https://ec.europa.eu/info/funding-tenders/opportunities/portal/screen/opportunities/topic-details/dt-spire-06-2019
25. HYPERCOG. 2019. https://www.hypercog.eu
26. Huertos F, Masenlle M, Chicote B, Ayuso M: Hyperconnected Architecture for High Cognitive Production Plants. 54th CIRP Conference on Manufacturing Systems, 2021
27. Huertos F, Chicote B, Masenlle M, Ayuso M: A Novel Architecture for Cyber-Physical Production Systems in Industry 4.0. CASE 2021: 645-650
28. Brandenburger J, Schirm C, Melcher J, Hancke E, Vannucci M, Colla V, Cateni S, Sellami R, Dupont S, Majchrowski A, Arteaga A: Quality4.0 - Transparent product quality supervision in the age of Industry 4.0. CoRR abs/2011.06502 (2020)
29. COGNITWIN. 2019. https://www.cognitwin.eu
30. Abburu S, Berre A, Jacoby M, Roman D, Stojanovic L, Stojanovic N: COGNITWIN - Hybrid and Cognitive Digital Twins for the Process Industry. ICE/ITMC 2020: 1-8
31. Jacoby M, Jovicic B, Stojanovic L, Stojanovic N: An Approach for Realizing Hybrid Digital Twins Using Asset Administration Shells and Apache StreamPipes. Inf. 12(6): 217 (2021)
32. Jacoby M, Volz F, Weißenbacher C, Stojanovic L, Usländer T: An approach for Industrie 4.0-compliant and data-sovereign Digital Twins. Autom. 69(12): 1051-1061 (2021)